\documentclass[floatfix]{revtex4}

\usepackage{epsfig}
\usepackage{graphicx}
\begin{document}
\bibliographystyle{revtex}


\title{Extended Gauge Sectors at Multi-TeV $e^+e^-$ Colliders}

\author{Stephen Godfrey}
\email[]{godfrey@physics.carleton.ca}
\affiliation{Ottawa-Carleton Institute for Physics \\
Department of Physics, Carleton University, Ottawa, Canada K1S 5B6}

\date{\today}

\begin{abstract}
We give a brief overview of searches at TeV $e^+e^-$ colliders
for new particles that arise in models with extended gauge sectors.  
We concentrate on some 
recent developments in $W'$ searches and on leptoquark searches.  
We also briefly mention $Z'$ searches.
\end{abstract}

\maketitle


\section{Introduction}
There are many models that approximate the standard model (SM) at present 
collider energies but that have a much richer particle spectrum above 100~GeV.  
A class of models that extends the SM in a natural way are extended 
gauge sectors \cite{ghp,c-g,dnr,e6}.  
These come in different varieties; Grand Unified Theories (GUTS)
such as $E_6$ and $SO(10)$ (and their Supersymmetric versions)
in which the subgroups of 
the SM are embedded into one larger group, of which an interesting variant 
is the Left-Right symmetric model embedded in a GUT, 
$E_6/SO(10) \to SU(3)_c  \times SU(2)_L \times SU(2)_R \times U(1)$ \cite{e6}.  
There are also numerous non-GUT models of extended gauge sector in the 
literature;  the Un-unified model, horizontal gauge models, 
and topcolour, to name but a few.  The point is 
that numerous possibilities exist that give rise to rich 
phenomenologies.
To distinguish among the many possibilities and 
reveal the underlying theory we will need to elucidate the TeV 
particle spectrum.

There are many new types of particles; Extra  $Z'$'s 
and $W'$'s bosons \cite{c-g,leike,godfrey,riemann}, 
New fermions of various types \cite{dnr} such as mirror fermions 
which are $SU(2)_R$ doublets and $SU(2)_L$ singlets, vector 
fermions which are both $SU(2)_R$  and $SU(2)_L$ doublets, and $SU(2)$ 
singlets like massive neutrinos, and Leptoquarks,  Diquarks and 
Bileptons.  Finally, an extended Higgs sector is probably the most 
obvious extension of the SM.  Establishing any extension of the 
standard will require two steps; Discovery and Identification.  For the 
former it is necessary to identify a signal that can be distinguished 
from background.  For identification there are many tools that can be 
used --- cross sections, angular distributions, decay signatures such 
as widths and branching ratios, and polarization observables.

It will not be possible to discuss these topics in detail in this 
brief contribution so I will limit myself to brief comments on $W'$'s 
and $LQ$'s and refer the interested reader to more detailed accounts 
elsewhere \cite{ghp,c-g,dnr,e6,leike}.

\section{Extra Gauge Bosons}

Extra  gauge bosons, both charged ($W'$) and/or neutral ($Z'$),
are a feature of many models of physics beyond the Standard 
Model (SM) \cite{c-g,leike}. 
A few models that have received attention recently 
and that we consider are 
the Left-Right symmetric model (LRM) based on the gauge group 
$SU(2)_L \times SU(2)_R \times U(1)_{B-L}$,
which has right-handed charged currents and restores parity at high 
energy,
the Un-Unified model (UUM) 
based on the gauge group $SU(2)_q \times SU(2)_l 
\times U(1)_Y$ where the quarks and leptons each transform under their own 
$SU(2)$, 
a Third Family Model (3FM) 
based on the gauge group $SU(2)_h 
\times SU(2)_l \times U(1)_Y$ where the quarks and leptons of the third 
(heavy) family transform under a separate group 
and the KK model (KK) 
which contains the Kaluza-Klein excitations of the SM gauge bosons that
are a possible consequence of theories with large extra dimensions. We 
also consider a $W'$ with SM couplings (SSM).
In studying extra gauge bosons, we are 
interested in two issues;  the sensitivity to $Z'$ and $W'$ discovery and
the measurement of their couplings to determine 
the underlying theory.

\subsection{$Z'$ Search and Identification}

In $e^+e^-$ collisions,  the $Z'$ contributes to the cross section via an 
s-channel propagator.  For $\sqrt{s}=M_{Z'}$ the resonance cross 
section is rather large, several orders of magnitude above the SM 
cross section, and the existence of a $Z'$ would be obvious.  In this 
case detailed studies could be performed and the properties of the 
$Z'$ could be extracted to high precision. Off resonance, the $Z'$ 
propagator interferes with that of the $\gamma$ and SM $Z^0$ leading 
to deviations from the SM expectations.  By combining the numerous 
available observables, a sensitivity to $Z'$'s with masses several 
times $\sqrt{s}$ can be achieved.  Depending on the mass of the $Z'$ 
one can also constrain its properties, such as it's couplings to 
different fermions.  Numerous studies exist in the literature 
\cite{c-g,leike,godfrey,riemann} and as 
we have nothing new to add at this point, we will not discuss $Z'$'s 
further.  

\subsection{$W'$ Search and Identification}

Until recently, $W'$ searches in the literature consisted of the $W$ 
pair production process $e^+e^-\to W_R^+ W_R^-$ and single $W$ 
production in $e^\pm \gamma \to W_R^\pm N$.  In both cases the
search limits are close to the kinematic limit which is
$\sim \sqrt{s}/2$ in the former case and $\sqrt{s_{e\gamma}}$ 
in the latter case.   
The exact limits will depend on the parameters of the model. 

Recently, it has  been demonstrated that limits greater than 
$\sqrt{s}$ can be achieved by considering 
the interference effects from t-channel $W'$ exchange 
 in the processes $e^+e^-\to \nu\bar{\nu}\gamma$ 
and $e\gamma \to \nu q +X$.  Here we give a brief summary of recent 
work on the subject and
refer the interested reader to the more detailed presentations of 
the process $e^+e^- \to \nu \bar{\nu} \gamma$ in Ref. \cite{wp1} 
and of the process $e \gamma \to \nu q +X$ in Ref.  \cite{wp2}.  

In the process $e^+e^-\to \nu\bar{\nu}\gamma$ \cite{wp1}, the signal is an 
energetic photon. The 
kinematic variables of interest are the photon's energy, $E_\gamma$, 
and its angle relative to the incident electron, $\theta_\gamma$, both 
defined in the $e^+e^-$ centre-of-mass frame.  To take into account 
finite detector acceptance we imposed the constraints on the kinematic 
variables: $E_\gamma \geq 10$~GeV and $10^o \leq \theta_\gamma \leq 
170^o$.
Several backgrounds should be taken into account.
The most dangerous is radiative Bhabba-scatter with the $e^+$ and 
$e^-$ go undetected down the beam.  
This can be eliminated by restricting the photon's transverse 
momentum to $p_T^\gamma > \sqrt{s}\sin\theta_\gamma \sin\theta_v 
/(\sin\theta_\gamma +\sin\theta_v )$ where $\theta_v=25$~mrad and is 
the minimum angle to which the veto detectors may observe electrons or 
positrons.  There are also higher order backgrounds which cannot be 
suppressed, such as $e^+e^- \to \nu\bar{\nu} \nu' \bar{\nu}'\gamma$, 
which must be 
included in the SM cross section but are fully calculable.  The 
low $E_\gamma$ region is most sensitive to the $Z'$.  The 
best limits are obtained by implementing a
kinematic cut on $E_\gamma$
to eliminate the radiative return to the SM $Z^0$-pole.  
The statistical significance can be increased by binning the $E_\gamma$
distribution  and calculating the $\chi^2$. The limits 
obtained by binning the $E_\gamma$
distribution  and calculating the $\chi^2$
with and without a 2\% systematic error added 
in quadrature with the statistical error are given in Table I.  
The limits were obtained using $e^-_L$ except for the LRM where $e^-_R$ 
was used.  
In general, the limits are highly model and machine dependent.

\begin{table}[t]
\caption{$W'$ discovery limits in TeV.  For $e\gamma\to \nu q +X$
the backscattered  laser photon spectrum was used.}
\centerline{
\begin{tabular}{lllllllll}
\hline
&\multicolumn{4}{c}{$\sqrt{s}=0.5$ TeV, $L_{int}=500$ fb$^{-1}$} &
\multicolumn{4}{c}{$\sqrt{s}=1$ TeV, $L_{int}=500$ fb$^{-1}$}\\
&\multicolumn{2}{c}{$e^+e^-\rightarrow \nu\bar\nu\gamma$} 
&\multicolumn{2}{c}{$e\gamma\rightarrow \nu q + X$} &
 \multicolumn{2}{c}{$e^+e^-\rightarrow \nu\bar\nu\gamma$}
&\multicolumn{2}{c}{$e\gamma\rightarrow \nu q + X$}\\
Model &no syst.&syst.&no syst&syst.&no syst.&syst.&no syst.&syst.\\ \hline
SSM $W'$  & 4.3 & 1.7 & 4.1 & 2.6 & 5.3 & 2.2 & 5.8 & 4.2 \\
LRM       & 1.2 & 0.6 & 0.8 & 0.6 & 1.6 & 1.1 & 1.2 & 1.1 \\
UUM       & 2.1 & 0.6 & 4.1 & 2.6 & 2.5 & 1.1 & 5.8 & 4.2 \\
3FM	  & 2.3 & 0.8 & 3.1 & 1.9 & 2.7 & 1.1 & 4.4 & 3.1 \\ 
KK        & 4.6 & 1.8 & 5.7 & 3.6 & 5.8 & 2.2 & 8.3 & 6.0 \\ \hline
\end{tabular}}
\end{table}

$W'$ couplings can also be constrainted using 
$e^+e^- \to \nu\bar{\nu} \gamma$.  
If a  $W'$ exists such that $M_{W'}$ is much lower than the search limit, 
we would 
expect it to be discovered at the LHC in which case we would want to 
measure its couplings. In Fig. 1 we show constraints for $W'$ 
couplings that could be obtained using $e^+e^- \to \nu\bar{\nu} \gamma$.

\begin{figure}
\includegraphics[width=1.85in]{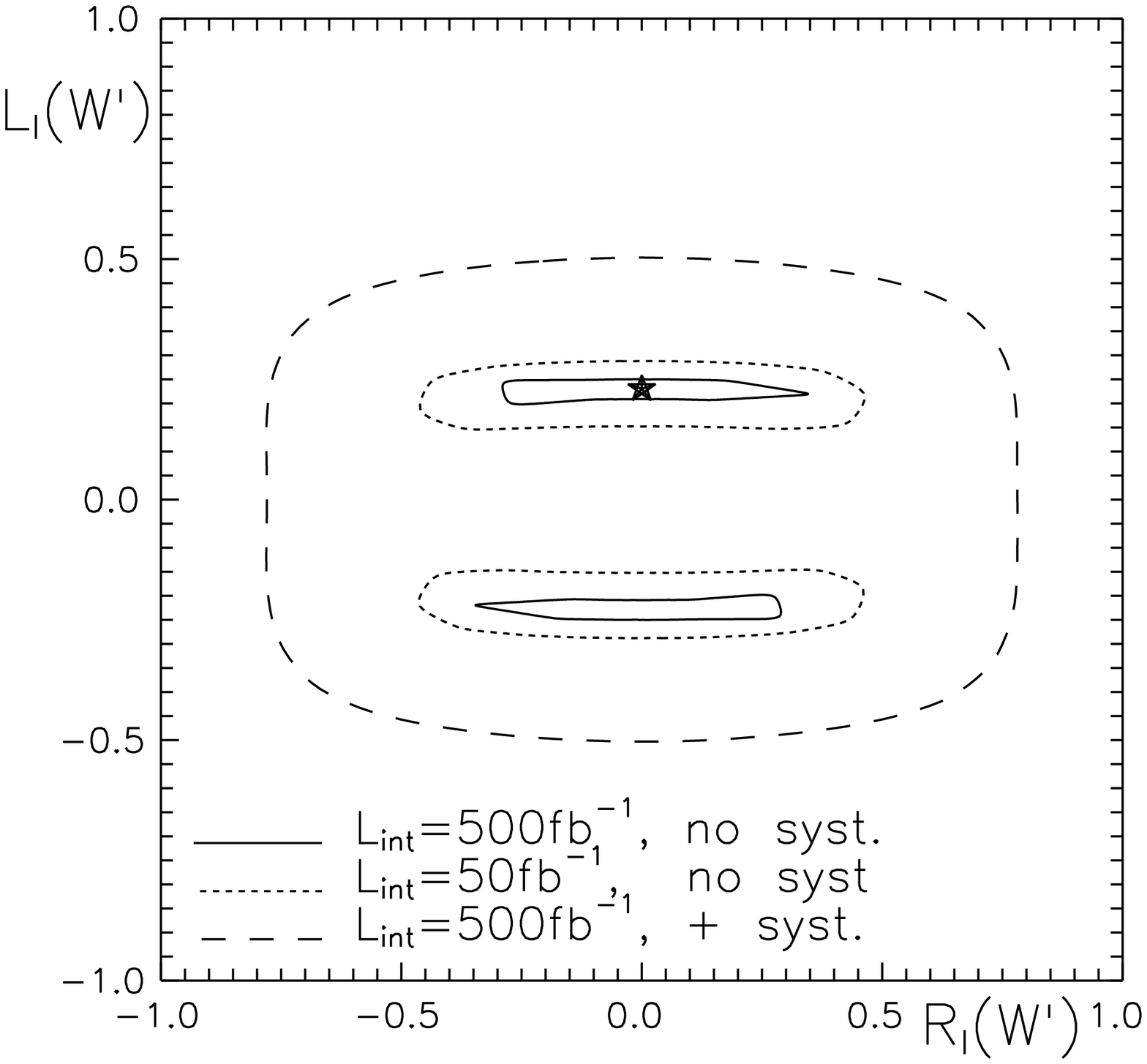} $\qquad\qquad$
\includegraphics[width=1.85in]{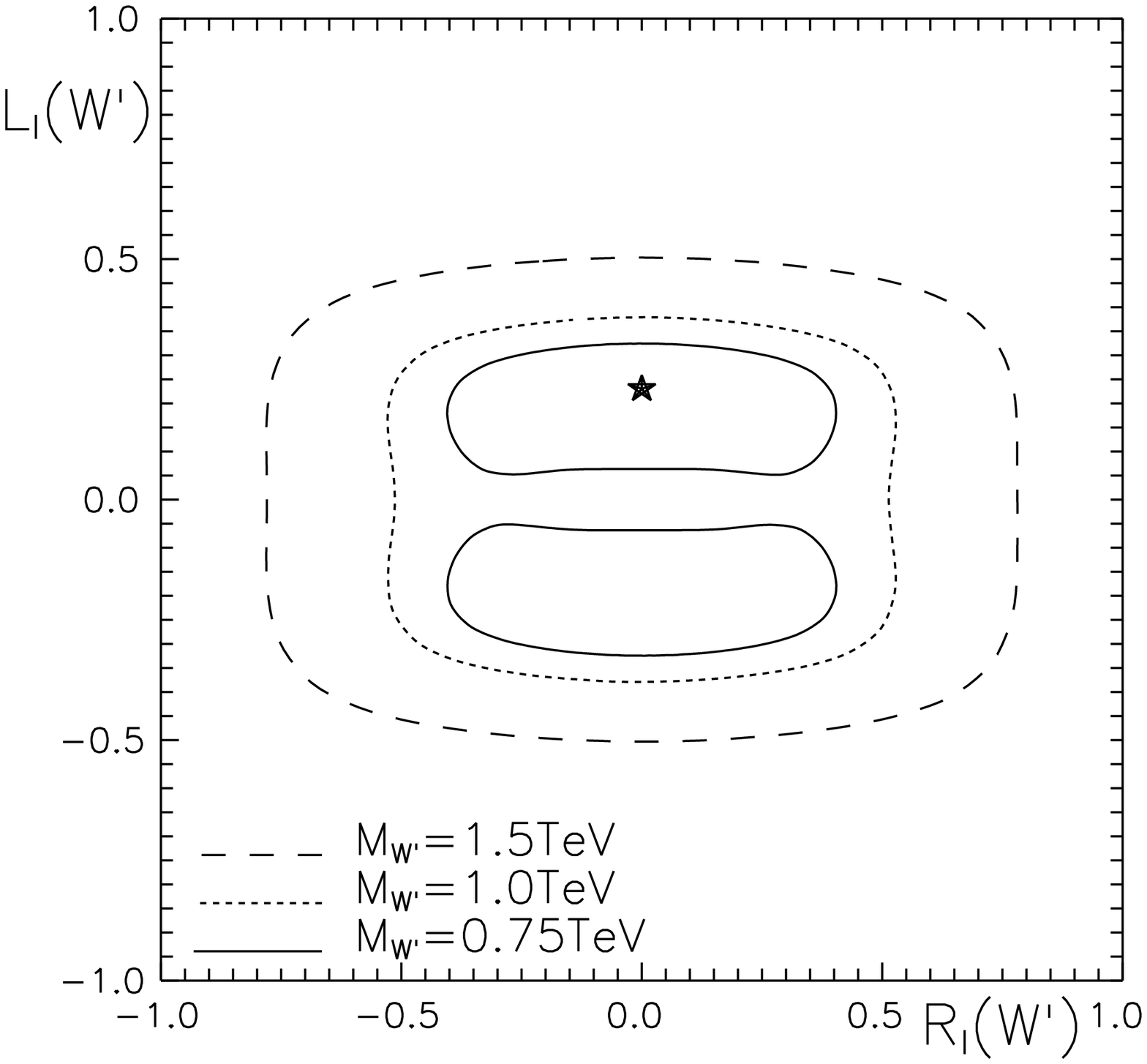}%
\caption{95\% C.L. constraints on
$L_f(W)  =  C_L^{W_i}g/(2\sqrt{2})$  and similarly for $R_f(W)$
based on $\sigma$ and $A_{LR}$ for  $\sqrt{s}=0.5$~TeV. 
We take 90\% electron and, where indicated, 60\% positron polarization.
The couplings of the assumed model, SSM $W'$, is indicated with a star.
(a) $M_{W'}=1.5$~TeV for integrated luminosities of 
$L_{\rm int}=50$ and 500~fb$^{-1}$ with and without a 2\% systematic 
error added in quadrature with the statistical error. 
(b) Limits for $M_{W'}=0.75$, 1.0, and 1.5~TeV  taking 
$L_{\rm int}=500$~fb$^{-1}$ .    
A systematic error of 2\% (1\%) is included for $\sigma$ ($A_{LR}$).
}
\label{Fig1}
\end{figure}

The processs $e^+e^- \to \nu\bar{\nu} \gamma$ is also 
sensitive to $Z'$'s.  Although this process 
is not as sensitive as other final 
states to $Z'$'s, it offers a means of measuring the 
$Z'-\nu\bar{\nu}$ coupling which could be useful in determining the 
$Z'$'s origin \cite{wp1}.

The sensitivity of the process $e\gamma \to \nu q +X$ to $W'$'s
was also studied \cite{wp2} where the $W'$'s enter via t-channel exchange. 
We considered 
two cases; the backscattered laser case and the Weizacker-Williams 
photon distribution which applies to $e^+e^-\to e^+ \nu q +X$. 
Starting with the subprocess $e\gamma \to 
\nu q\bar{q}$ the $W'$ contributions can be enhanced 
by imposing the kinematic cut that either the $q$ or $\bar{q}$ is 
collinear to the beam axis.  In this kinematic region the process 
$e\gamma\to \nu q\bar{q}$ is approximated quite well by the simpler 
process $e q\to \nu q'$ where the quark is described by the quark 
parton content of the photon, the so-called resolved photon 
approximation.  
The process $eq\to \nu q'$ was used to obtain 
limits as it is computationally much faster and free of sensitivity to 
the quark masses and collinearity cuts.

As usual, kinematic cuts were included to reflect finite detector 
acceptance; 
the angle of the outgoing $q(\bar{q})$ was restricted
to the range $10^o \leq 
\theta_{q(\bar{q})} \leq 170^o$.  We included $u$, $d$, $s$, and 
$c$-quark contributions and used the leading order GRV 
distributions\cite{grv}.  
The search limits are fairly insensitive to the specific choice of 
distribution.  The kinematic variable most sensitive to $W'$
is the $p_{T_q}$ distribution.  
$\sigma_R$ was only found to be sensitive 
to the LR model but even small $e_L$  pollution swamps $\sigma_R$.  

As always, it is necessary to consider and 
eliminate serious backgrounds.  The dominant backgrounds are 
two jet final states comprised of
$\gamma\gamma \to q\bar{q}$, the once resolved 
reactions $\gamma g \to q\bar{q}$ and $\gamma q \to g q$, and the 
twice resolved reactions $gg \to q\bar{q}$, 
$q\bar{q} \to q\bar{q}$, $qg \to qg$, $gg \to gg$, $qq \to qq$ ...,
where one of the jets goes down the beam pipe and is not observed. 
These backgrounds can be effectively eliminated by imposing the
constraint $p_{T_q} > 40, 75, 100$~GeV for $\sqrt{s}=0.5, 1.0,$ and 
1.5~TeV, respectively.
Discovery limits were obtained by binning the 
$p_{T_q}$ distribution and calculating the $\chi^2$ for an assumed 
integrated luminosity.  As before, a 2\%  systematic error was 
included in quadrature with the statistical error.  The discovery 
limits using the backscattered laser spectrum are given in Table I. 
In all cases, the limits from the backscattered laser are better than 
those from the Weizsacker Williams process due to the harder photon 
spectrum of the former. Again, the limits are given for unpolarized beams. 
Electron beam polarization was not found to yield significantly improved 
results for this process.

The limits from the process $e\gamma\to \nu q+X$ compare favourably with 
those from $e^+e^- \to \nu \bar{\nu} \gamma$ in all models other than the 
LRM. In that case, the limits are comparable for the two processes. The LHC 
is expected to detect $W'$'s up to a mass of about 5.9 TeV \cite{lhc}, 
although that number is also highly model dependent. Hence the process 
$e\gamma\to \nu q+X$ shows promise even compared with the reach of the LHC. 
Therefore, a more detailed consideration of the exact process, rather than 
the resolved photon approximation given here, and including radiative 
corrections, is motivated.

In Fig. 2 we show the contraints on $W'$ couplings that can be 
obtained from the process $e\gamma \to \nu q +X$.
We assume that a $W'$ has been discovered elsewhere and its mass is 
known.  The contours are based on the ${p_T}_q$ distribution.

\begin{figure}
\includegraphics[width=1.85in]{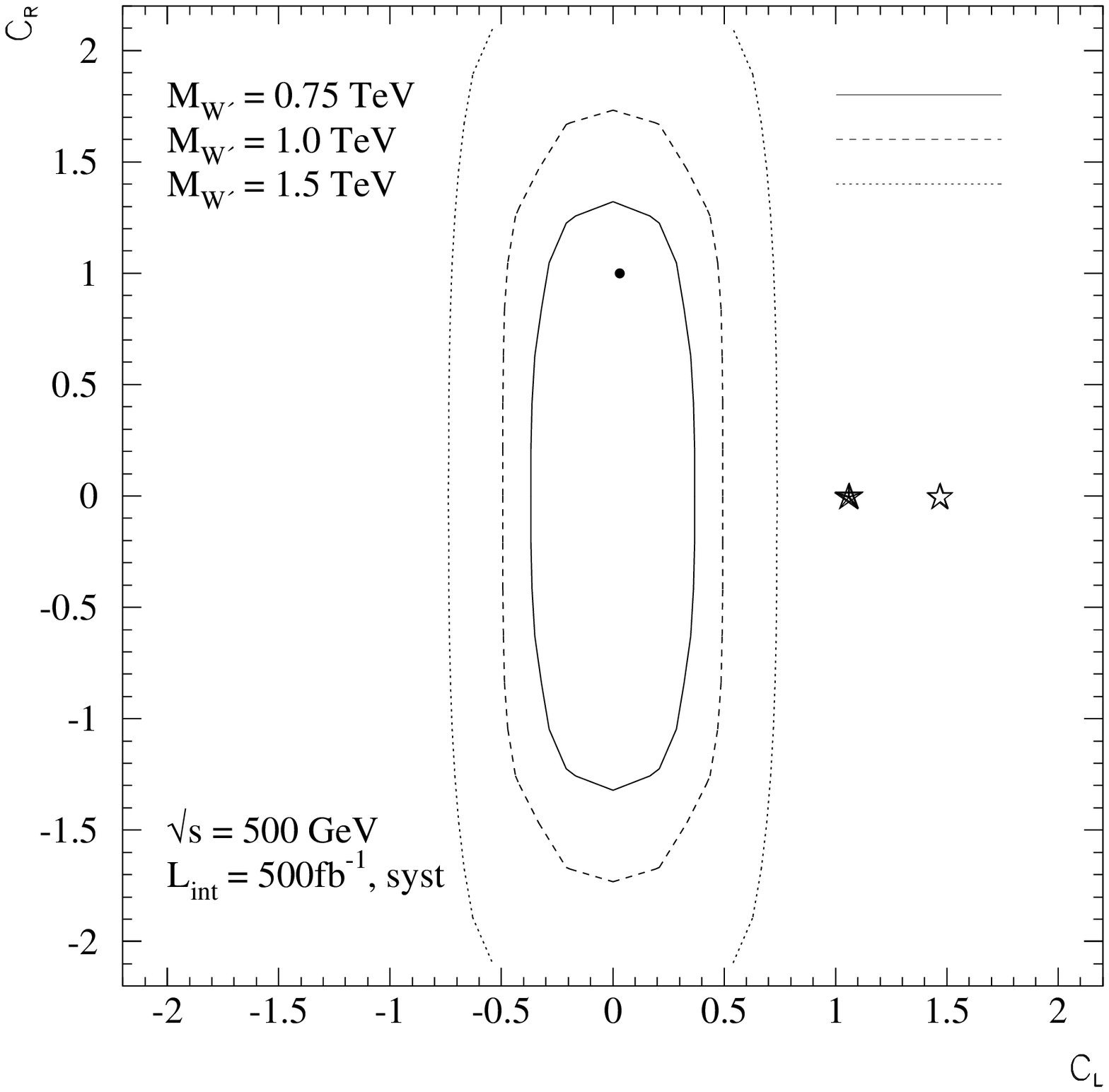} $\qquad\qquad$
\includegraphics[width=1.85in]{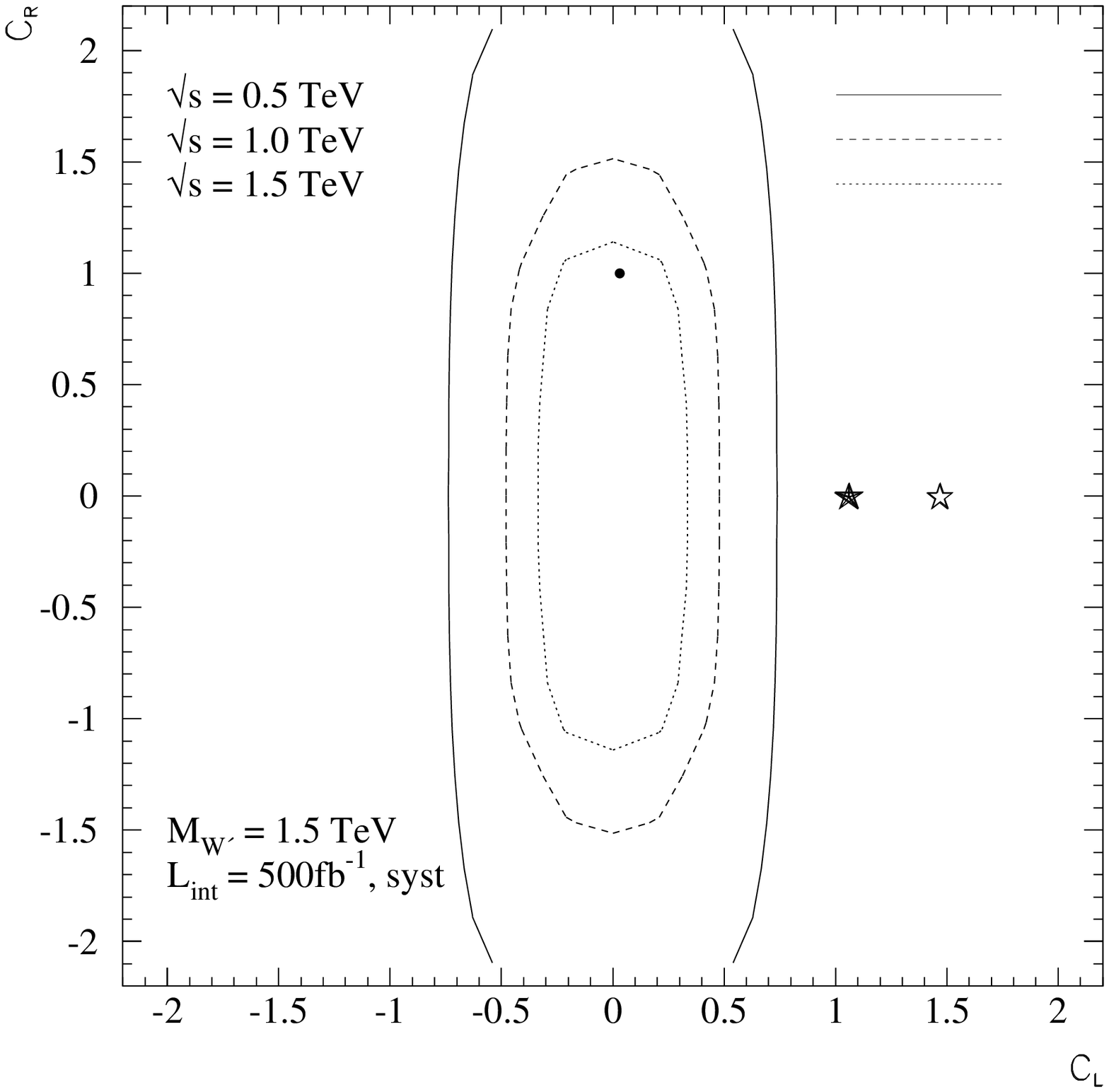}%
\caption{95\% C.L. constraints on $W'$ couplings 
from the process $e\gamma \to \nu q +X$ 
with a backscattered laser spectrum assuming the SM and 
using the $d\sigma/dp_{T_q}$ with a 2\% systematic 
error.  We have taken
$C_{L(R)}^e=C_{L(R)}^q$ which is satisfied in many models.
The SSM, LRM and the KK model are indicated 
by a full star, a dot and an open star, respectively.
(a) For different values of $M_{W'}$ with $\sqrt{s}=500$~GeV. 
(b) For different values of $\sqrt{s}$ with $M_{W'}=1.5$~TeV.
}
\label{Fig2}
\end{figure}

\section{Leptoquarks}

Leptoquarks appear in many theories, from GUTS to technicolour to 
composite models.  
Leptoquarks are colour triplets or antitriplets that carry both baryon 
and lepton quantum numbers and can have either spin 0 or 1.  According 
to the Buchmuller-Ruckl-Wyler classification there are 10 distinct types 
of LQ's.  At the LHC, with $L_{\rm int}=100$~fb$^{-1}$, LQ's can be 
discovered up to $M_{LQ}=1.4$~TeV and 2.2~TeV for scalar and vector 
LQ's respectively.  

We consider LQ production via the quark content of the photon in 
$e\gamma$ collisions \cite{lq}.  
The quark fuses with the electron to produce a 
LQ and the $\sigma(eq\to LQ)$ cross section is 
convoluted with the quark distribution inside the photon which is 
subsequently convoluted with the photon distribution function.   
Cross sections for LQ production for the $e\gamma$ case using a 
backscattered laser and for the $e^+e^-$ case with Weizacker-Williams 
photons are shown in Fig. 3. In addition to direct production one can also 
obtain indirect limits via t-channel LQ exchange in $e^+e^-\to 
q\bar{q}$ \cite{hr-lq}.

\begin{figure}
\includegraphics[width=7.3cm]{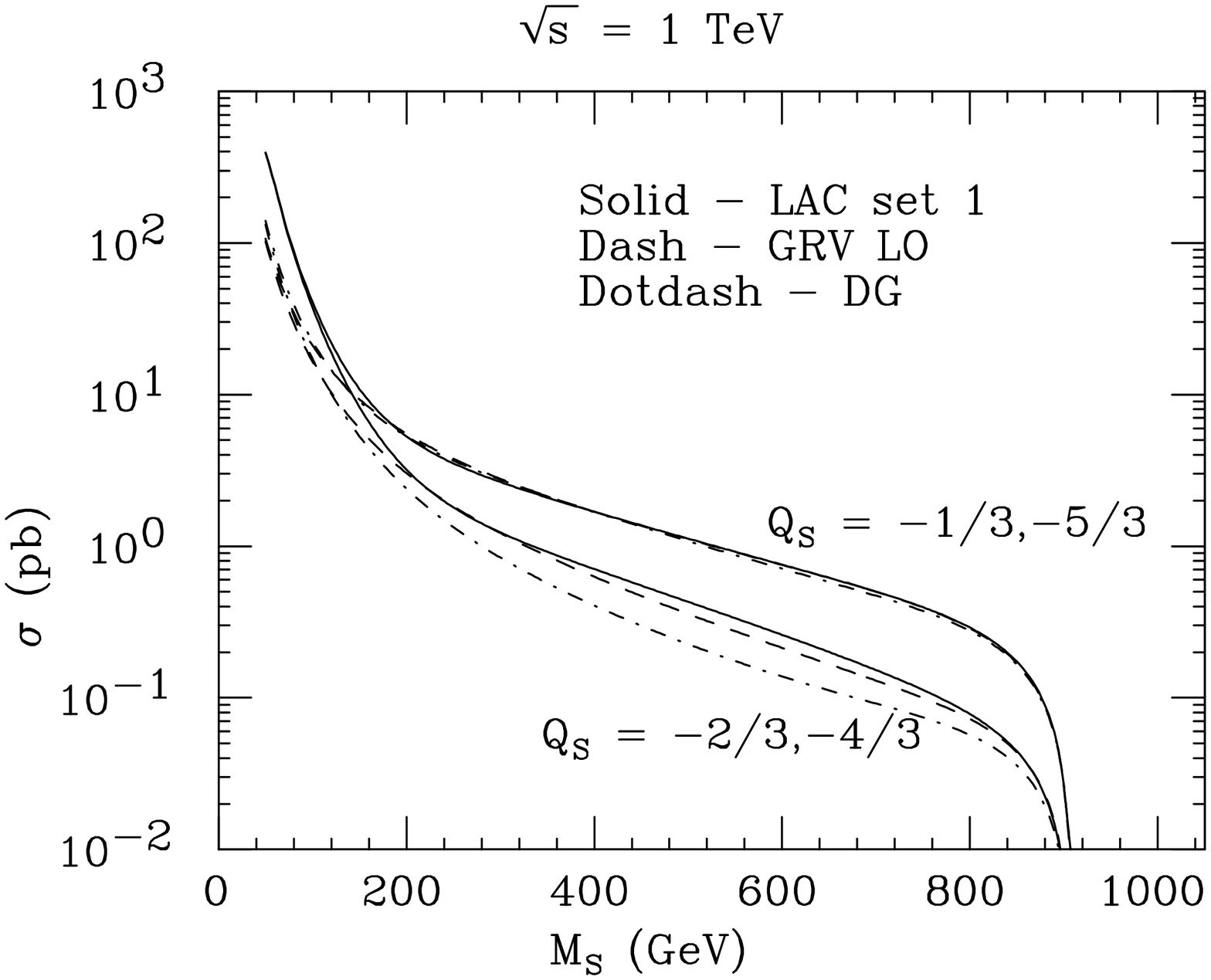} $\qquad\qquad$
\includegraphics[width=7.3cm]{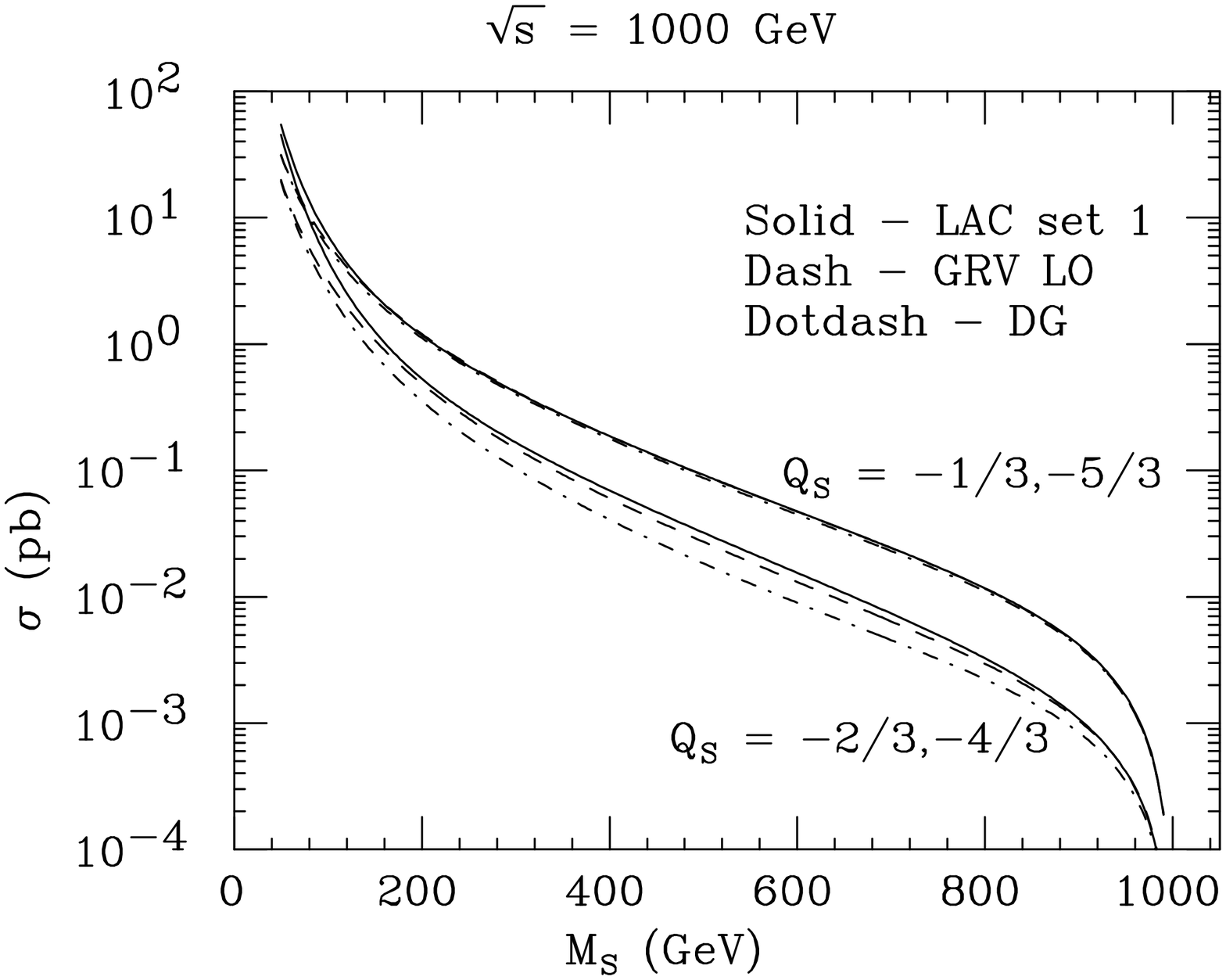}%
\caption{The cross sedtions for leptoquark production due to resolved 
photon contributions in $e\gamma$ collisions from 
a $\sqrt{s}=1$~TeV $e^+e^-$ collider.
(a)  The photon beam is due 
to laser backscattering in the $e^+e^-$ collider.  
(b) The photon distribution is given by the Weizs\"acker-Williams 
effective photon distribution.
}
\label{Fig3}
\end{figure}
 
For the luminosities envisaged at future $e^+e^-$ colliders such as 
the LC and CLIC, LQ's can be discovered almost up to the kinematic 
limit.  Discovery limits are given in Table II.
The limits for $q=-1/3$ and $-5/3$ LQ's are slightly  larger than for
$q=-4/3$ and $-2/3$ LQ's because of the higher u-type quark content of 
the photon compared to the d-type quark content.  The vector LQ's have 
slightly higher limits compared to scalar LQ's because of vector vs 
scalar couplings to fermions.  Finally, it turns out that for some 
cases $e^+e^-$ colliders yield higher limits than $e\gamma$ colliders 
because for $e\gamma$ colliders the maximum $\sqrt{s_{e\gamma}}$ 
is less than that of an $e^+e^-$ collider and 
the high luminosities more than 
compensate for the  softer photon spectrum.  It should be noted that 
this search strategy has been employed by the OPAL \cite{opal} and 
DELPHI \cite{delphi} collaborations to obtain LQ limits.

\begin{table}
\caption{Leptoquark discovery limits for $e^+e^-$ and $e\gamma$ 
colliders.  The discovery limits are based on the production of 100 
LQ's for the centre of mass energies and integrated luminosities given 
in columns one and two.  The results were obtained using the GRV 
distribution functions.}
\begin{tabular}{llcccccccc}
\hline
& & \multicolumn{4}{c}{$e^+e^-$ Colliders} 
	& \multicolumn{4}{c}{$e\gamma$ Colliders}\\ \hline
$\sqrt{s}$  & $L$  & \multicolumn{2}{c}{Scalar} & 
\multicolumn{2}{c}{Vector}  
& \multicolumn{2}{c}{Scalar} & \multicolumn{2}{c}{Vector} \\ \hline 
(TeV) & (fb$^{-1}$) & (-1/3, -5/3) & (-4/3, -2/3) & (-1/3, -5/3)
 & (-4/3, -2/3) 
& (-1/3, -5/3) & (-4/3, -2/3) & (-1/3, -5/3) & (-4/3, -2/3) \\ \hline 
 0.5  & 50 & 490 & 470 & 490 & 480 & 450 & 450 & 450 & 440  \\
 1.0  & 200 & 980 & 940 & 980 & 970 & 900 & 900 & 910 & 910 \\
 1.5 & 200 & 1440 & 1340 & 1470 & 1410 & 1360 & 1360 & 1360 & 1360 \\
 5.0 & 1000 & 4700 & 4200 & 4800 & 4500 & 4500 & 4400 & 4500 & 4500 \\ 
\hline
\end{tabular}
\end{table}

If leptoquarks were discovered their production in $e\gamma$ 
collisions could be used to identify them which would be crucial for 
determining their theoretical origin.  Their decay distributions can 
be used to determine whether they are scalar or vector and electron 
polarization asymmetries can be used to determine the chirality of 
their couplings to fermions.  Finally, the cross sections of the 
different LQ's are different enough that they can be separated on this 
basis to relatively high masses.

\section{Summary and Outlook}

Extended gauge sectors give rise to a very rich phenomenology.  In 
this brief report we only touched the surface, neglecting many 
important topics 
such as extra fermions, bileptons and diquarks.  

A considerable body of work exists on $Z'$'s and a large effort 
continues to be applied to the topic.  The exclusion limits that can 
be obtained at TeV linear colliders are similar to, or exceed, the 
discovery limits of the LHC. If a $Z'$ is discovered at the LHC, the LC 
would be an important tool for its identification.  
Recently,  indirect effects of $W'$'s in $e^+e^-$ and $e\gamma$ 
collisions have also been studied and it was found that measurements 
at $e^+e^-$ colliders are sensitive to $W'$ bosons considerably higher 
in mass than their centre-of-mass energy.
For some models, 
the reach that can be obtained at a LC is competitive with that of the LHC,
particularly in the case of the process $e\gamma\to \nu q +X$. 
If a $W'$ were discovered at the LHC the LC would be an 
important tool in measuring their properties.  However, 
there have only been a few such studies and there is room for new 
ideas.   The studies of both $Z'$'s and $W'$'s need to be updated to
include the  
energies relevant to CLIC and to take into account the higher 
luminosities now envisaged.  
Leptoquarks could be discovered for masses up to almost $\sqrt{s}$. 
If LQ's are 
discovered, $e\gamma$ could make significant contributions to their 
understanding.  

To conclude,  high energy $e^+e^-$ colliders have considerable 
potential for the discovery of the particles expected in extended 
gauge theories.  But more than that, they could play a crucial role in 
measuring the properties of new particles, and hence unravel the 
underlying theory.   

%
%
%
%



\end{document}